\documentclass[journal,draftcls,onecolumn,12pt,twoside]{IEEEtran}
\normalsize
\usepackage[colorlinks,
linkcolor=black,
anchorcolor=black,
citecolor=black,
]{hyperref}
\usepackage{diagbox}
\usepackage[english]{babel}
\usepackage{latexsym}
\usepackage{varioref}
\usepackage{array}
\usepackage{hhline}
\usepackage{dcolumn}
\usepackage{tabularx}
\usepackage{algorithm}
\usepackage{algorithmic}
\usepackage{calc}
\usepackage{epsfig}
\usepackage[usenames]{color}
\usepackage{amsmath,amssymb,bigdelim}
\usepackage{t1enc}
\usepackage{pstricks,pst-node,psfrag,pst-plot}
\usepackage{graphicx,parskip,amsbsy}
\usepackage{boxedminipage}
\usepackage{indentfirst}
\usepackage{booktabs}
\usepackage[noadjust]{cite}
\usepackage{caption}
\usepackage{multirow}
\usepackage{subfigure}
\usepackage[usenames]{color}
\usepackage{clrscode}
\newcommand{\rem}[1]{}

\providecommand{\texlivekeywords}[1]{\textbf{\textit{Index terms---}} #1}
\begin{document}
	
	\pagestyle{plain}
	\title{{ A Complexity-Efficient} High Resolution Propagation Parameter Estimation Algorithm for Ultra-Wideband Large-Scale Uniform Circular Array}
	\author{ Xuesong Cai and Wei Fan 
		\thanks{X. Cai and W. Fan are with the APMS section at the Department of Electronic Systems, Falculty of Engineering and Science, Aalborg
University, Aalborg 9220, Denmark (e-mail: xuc@es.aau.dk; wfa@es.aau.dk).


}
%
%
%
%
	}

\markboth{IEEE Transactions on Communications}%
{Submitted paper}

	\maketitle \thispagestyle{plain}
\begin{abstract}

Millimeter wave (mm-wave) communication with large-scale antenna array configuration is seen as the key enabler of the next generation communication systems. Accurate knowledge of the mm-wave propagation channels is fundamental and essential. 
In this contribution, a novel complexity-efficient high resolution parameter estimation (HRPE) algorithm is proposed for the mm-wave channel with large-scale uniform circular array (UCA) applied. The proposed algorithm is able to obtain the high-resolution estimation results of the spherical channel propagation parameters. The prior channel information in the delay domain, i.e., the delay trajectories of individual propagation paths observed across the array elements, is exploited, by combining the high-resolution estimation principle and the phase mode excitation technique. Fast initializations, effective interference cancellations and reduced searching spaces achieved by the proposed schemes significantly decrease the algorithm complexity. Furthermore, the channel spatial non-stationarity across the array elements is considered for the first time in the literature for propagation parameter estimation, which is beneficial to obtain more realistic results as well as to decrease the complexity. A mm-wave measurement campaign at the frequency band of 28-30\,GHz using a large-scale UCA is exploited to demonstrate and validate the proposed HRPE algorithm.

\end{abstract}
\texlivekeywords{
Millimeter-wave, ultra-wideband, large-scale uniform circular array, spatial non-stationarity and channel parameter estimation.
}
\IEEEpeerreviewmaketitle
\section{Introduction}
While the next generation communication system (5G) is still on its ascendant, millimeter wave (mm-wave) communication has been seen as the key enabling component due to the vast amount of available spectrum \cite{6515173,7894280,6732923}. However, the air attenuation, small antenna aperture etc. would result in severe power loss \cite{6894453} for the mm-wave propagations, compared to the sub-6\,GHz frequency bands with rich multipath components (MPCs) \cite{8240983}. Nevertheless, the mm-wavelength makes it practical to pack massive antennas \cite{8114238} in a small area to form large scale antenna arrays. Beamforming \cite{7731243} and beam-selection \cite{7894280} techniques can be exploited to compensate the power loss and to enhance the spectrum efficiency through spatial multiplexing. Furthermore, applications such as localization, tracking and surrounding environment reconstruction \cite{7426565} are promising, e.g., to help vision-disabled people.

To enable the advanced 5G mm-wave techniques and applications, accurate and realistic channel models are fundamental and essential. The establishment of effective mm-wave channel models relies on the comprehensive channel measurements and the channel characteristics extracted from the measured data. In the mm-wave propagations with large scale antenna array configurations, the assumptions applied for the previous generation communication systems (e.g. the Long Term Evolution system) are violated. \textit{i)} The narrowband assumption \cite{526899} is invalid due to the ultra-wide system bandwidth up to several GHz; \textit{ii)} The two dimensional (2D) propagation assumption \cite{753729} was usually assumed in pervious communication systems since the base station was high above, and the propagation distance was large. However, in the mm-wave frequency bands, the propagation distance is generally limited due to the high power loss. It is not practical anymore to assume the 2D propagation. That is, the elevation angels of MPCs \cite{7572761} should be considered; \textit{iii)} The mm-wavelength and large array aperture can result in the Fraunhofer far-field distance up to tens or hundreds of meters, which is significantly increased compared to the previous systems. This necessitats the spherical wave propagation \cite{7981398} model; \textit{iv)} It is possible that the gain (not just phases) of propagation paths evolve \cite{7172496,7817797} across the array elements. The so-called spatial non-stationarity in path gain
must be considered as well. Thus the mm-wave channel models must provide realistic channel characteristics in multiple parameter domains, i.e., in delay, azimuth, elevation, source distance and amplitude domains. Consequently, accurate, comprehensive yet still complexity-efficient channel parameter estimation algorithms are vital to rapidly promote the 5G progress.

Among the many mm-wave channel spatial profile measurements, the most frequently applied measurement approach is to either mechanically rotate a horn antenna or mechanically move an omnidirectional antenna to form virtual arrays \cite{7913702}. The vector network analyzer (VNA) is usually used due to its flexibility and easy-configuration. Various estimation algorithms have been proposed for the mm-wave channel estimation based on the empirically measured data. Basically, these algorithms can be classified into four categories, i.e., spectra based approaches, subspace based approaches, sparsity recovery approaches and maximum likelihood approaches as follows. \textit{i)} For the horn antenna measurements, the measured channels at different steering directions constructed the joint delay-angle power spectra, e.g., as done in \cite{8103059,8094309,7400962,7109864}. The complexity is of the algorithm is low, while the resulted channel characteristics show dependency on the horn antenna pattern. In \cite{7343290}, the fast fourier transform (FFT) was applied for the extremely large antenna array to distinguishing MPCs.
A frequency invariant beamformer was proposed in \cite{7523340} and extended in \cite{ZhangArxiv} for a uniform circular array (UCA) to obtain the delays and azimuths of MPCs in the 3D propagation scenarios.
\textit{ii)} In \cite{7938435}, the multiple signal classification (MUSIC) principle was applied to jointly estimating the delay and azimuth in a 2D mm-wave propagation scenario. Unitary estimation of signal parameter via rotational invariance techniques (ESPRIT) was also exploited in \cite{8313174} for the 3D mm-wave propagation scenarios with plane wave assumption. \textit{iii)} Based on the sparsity assumption of mm-wave channels, the authors in \cite{7390019,8269069,7227017,8122055} recovered the channels and obtained the channel estimations by exploiting the compressive sensing or optimization techniques. \textit{iv)} 
The Space Alternating Generalized Expectation-maximization (SAGE) algorithm was applied for the mm-wave channel estimation in \cite{5956639,aaltosage,7817797,7501567,7981398}. Both the 3D plane-wave propagation scenarios \cite{5956639,aaltosage,7817797} and the 3D spherical propagation scenarios \cite{7501567,7981398} were concerned. However, the fundamental assumption of SAGE is that the received signal contains multiple components that follow orthogonal stochastic measures (OSM) \cite{324732,7981398}. The violation \cite{7744514} of the OSM assumption caused by 3D spherical propagation could deteriorate the SAGE performance significantly, e.g., leading to the non-convergence. A maximum likelihood estimator (MLE) based on the expectation-maximization (EM) principle was proposed in \cite{yilin} aimed for the 3D spherical propagation scenario. The complexity is still considerably high with the 4D parameter searching, though a coarse-to-fine search strategy was applied.

Although efforts have been made to characterize the mm-wave propagation channels, a high resolution propagation parameter estimation (HRPE) algorithm which is capable to obtain the high-resolution estimation results of all the spherical propagation parameters (i.e. azimuths, elevations, delays, source distances and amplitudes of MPCs) yet with low computation complexity is still missing in the literature. The existing algorithms are either deficient in obtaining the high-resolution estimation results of all propagation parameters or with fatal computation complexity especially when considering the ultra-wide bandwidth and large-scale array aperture in mm-wave propagation. Moreover, as far as we are concerned, the realistic spatial non-stationarity across the array elements have never been considered for propagation parameter estimation, although it has been widely observed in the mm-wave channel modeling works in the literature, leading non-realistic estimation results and increasing complexity. To fill the above gaps, a complexity-efficient HRPE algorithm is proposed in this paper for the ultra-wideband large-scale UCA. The UCA is considered as it is capable to uniformly cover the whole 360$^\circ$ of azimuth compared to the linear arrays and support for phase mode operations. The main contributions and novelties of this paper include:
%

\begin{itemize}
\item The proposed algorithm exploits the high resolution estimation technique in the delay domain, the UCA phase mode excitation technique and the maximum-likelihood estimation principle. Fast initializations, effective interference cancellations and reduced searching spaces substantially decrease the computation complexity while maintaining the ability to gain the whole high-resolution estimation results of the 3D spherical mm-wave propagation channel parameters.

\item The UCA phase mode excitation technique applied in the 3D spherical mm-wave propagation scenarios is investigated through simulations, in terms of its ability to obtain the fast initializations of delays and azimuths.

\item Individual propagation paths are firstly identified in the delay domain, so that the fast initializations and effective interference cancellations can be achieved for the spherical propagation parameter estimation. Furthermore, the spatial non-stationarity observed across the array elements is considered for the first time in the literature for the propagation parameter estimation, resulting in more realistic estimation results and significantly reduced complexity.

\item An ultra-wideband mm-wave channel measurement at the frequency band of 28-30\,GHz with large-scale (radius of 0.5\,m and 720 elements) UCA applied was conducted. The comparisons among the reference channel measured by using the horn antenna, the MLE (or EM) results obtained by applying 4D parameter searchings \cite{yilin} and the results obtained by using the proposed HRPE algorithm validate the algorithm performance.
\end{itemize}

The rest of the paper is structured as follows. Sect.\,\ref{signalmodel} elaborates the signal model. Sect.\,\ref{section:setup_and_environment} describes the measurement campaign and the observed channel characteristics.
Sect.\,\ref{algorithm} elaborates the proposed HRPE algorithm. The algorithm validation and the remarks are included in Sect.\,\ref{resultsremarks}. Finally, conclusions are given in Sect.\,\ref{section:conclusion}.

\section{Signal Model for The Uniform Circular Array\label{signalmodel}}

\begin{figure}
\begin{center}
\psfrag{a}[c][c][0.7]{$\theta_\ell$}
\psfrag{b}[l][l][0.7]{$\phi_p$}
\psfrag{c}[c][c][0.7]{UCA center}
\psfrag{x}[c][c][0.7]{(reference point)}
\psfrag{d}[c][c][0.7]{$\phi_\ell$}
\psfrag{e}[r][r][0.7]{Wave $\# \ell$}
\psfrag{f}[l][l][0.7]{Wave $\# \ell^\prime$ }
\psfrag{g}[r][r][0.7]{Wave $\# L$}
\psfrag{h}[c][c][0.7]{$d_\ell, \tau_\ell, \alpha_\ell$}
\psfrag{r}[c][c][0.7]{$r$}
\psfrag{R}[c][c][0.7]{Rx UCA}
\psfrag{T}[c][c][0.7]{Tx}
\psfrag{s}[c][c][0.7]{Scatterrer}
\includegraphics[width=0.5\textwidth]{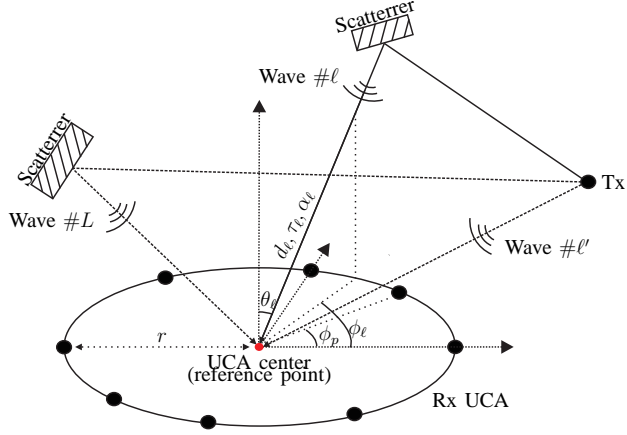}
\end{center}
\caption{Multipath prorogation with UCA.\label{fig:uca}}
\end{figure}

As illustrated in Fig.\ \ref{fig:uca}, the UCA has $P$ isotropic receiver antenna elements (Rxs) uniformly arranged on its perimeter with radius of $r$, i.e., the azimuth of the $p$th Rx is $\phi_p=\frac{2\pi p}{P}, p=[0, \cdots, P-1]$. 

With a large UCA aperture, the spherical wave signal model has to be assumed due to the violation of plane wave assumption. In the underlying channel model, a finite number of $L$ spherical waves are assumed to impinge into the UCA. Although an arbitrary reference point can be selected, we consider the UCA center as the reference point for the model conciseness. The frequency response $H_{\ell}(f)$ at the UCA center, which is contributed by the $\ell$th path (or wave) with parameter set $\Theta_\ell=[\tau_\ell, \phi_\ell, \theta_\ell, d_\ell, \alpha_\ell]$, reads
\begin{equation}
\begin{aligned}
H_{\ell}(f) = \alpha_\ell e^{-j 2\pi f \tau_\ell}
\end{aligned}
\end{equation}
where $f=[f_1,\cdots,f_K]$ is the vector containing the $K$ frequency points considered, $\Theta_\ell$ is the vector containing the parameters of the $\ell$th path, where $\tau_\ell$ is the propagation delay, $\phi_\ell$ and $\theta_\ell$ represent the azimuth and elevation angles, respectively, $d_\ell$ is the propagation distance between the UCA center and the last {source point during the propagation route}, and $\alpha_\ell$ denotes the complex amplitude. The propagation distance between the $p$th Rx and the last source point is then calculated as
\begin{equation}
\begin{aligned}
 d_{p,\ell} = \sqrt{{d_\ell}^2 + r^2 - 2rd_\ell\sin\theta_\ell\cos(\phi_\ell - \phi_p)}
\end{aligned}
\label{eq:distance}
\end{equation}
Due to the propagation distance difference between the $p$th Rx and the UCA center, the frequency response $H_{p,\ell}(f)$ contributed by the $\ell$th path at the $p$th Rx consequently reads
\begin{equation}
\begin{aligned}
H_{p,\ell}(f) & = \frac{d_\ell}{d_{p,\ell}} H_{\ell}(f) e^{j 2\pi f \Delta d_{p,\ell}/c}
\end{aligned}
\label{eq:elesage}
\end{equation}
where
\begin{equation}
\begin{aligned}
\Delta d_{p,\ell} & = d_\ell- d_{p,\ell}
\end{aligned}
\label{eq:distancediff}
\end{equation}
with $c$ the light speed, and $\frac{d_\ell}{d_{p,\ell}}$ denotes the amplitude change factor according to the Friis transmission equation.  
The array output contributed by the $\ell$th path then reads $H(p,f;\Theta_\ell) = [H_{0,\ell}, \cdots, H_{P-1,\ell}]^T$, where $[\cdot]^T$ denotes the transpose operation of the argument. $H(p,f;\Theta_\ell)$ is a complex-valued matrix with dimension of $P\times K$, i.e., $H(p,f;\Theta_\ell)\in \mathbb{C}^{P\times K} $.

The array output contributed by all the $L$ paths is formatted as
\begin{equation}
\begin{aligned}
Y(p,f;\Theta) = \sum_{\ell=1}^{L} H(p,f;\Theta_\ell) + n(p,f)
\end{aligned}
\end{equation}
where $n(p,f)\in \mathbb{C}^{P\times K}$ is the zero-mean white Gaussian noise with variance of $\sigma^2$, and $\Theta = [\Theta_1, \cdots, \Theta_L]$ is the vector containing all the channel parameters. The objective of various estimation algorithms is to obtain the estimation results of $\Theta$. In the sequel, a novel HRPE algorithm is proposed for the ultra-wideband large-scale UCA, which is demonstrated based on a measurement campaign where the realistic channel characteristics are well presented.  The proposed algorithm is capable to gain the estimation of $\Theta$.


\section{Measurement campaign and channel observations}\label{section:setup_and_environment}
\subsection{Measurement campaign}
In this section, a recently conducted measurement campaign for a UCA is introduced. The measurement scenario and the measurement setup including the transmitter (Tx), Rx, array aperture, frequency range, bandwidth etc. are described.

\begin{figure}
\begin{center}
\psfrag{K}[c][c][0.6]{Wall B}
\psfrag{j}[c][c][0.6]{Wall }
\psfrag{V}[c][c][0.6]{Wooden wall A}
\psfrag{L}[c][c][0.5]{$0.5$ m}
\psfrag{C}[c][c][0.6]{Corridor}
\psfrag{S}[c][c][0.6]{Window and metallic stairs}
\psfrag{H}[c][c][0.6]{Metallic heater}
\psfrag{d}[c][c][0.6]{$3.6$ m}
\psfrag{e}[c][c][0.6]{$4.1$ m}
\psfrag{g}[c][c][0.6]{$0.9$ m}
\psfrag{f}[c][c][0.6]{$5.0$ m}
\psfrag{n}[c][c][0.6]{$1.7$ m}
\psfrag{T}[c][c][0.6]{Tx}
\psfrag{b}[c][c][0.6]{Blackboard}
\psfrag{c}[c][c][0.6]{$3$ m}
\psfrag{R}[c][c][0.6]{Starting point of clockwise virtual Rx positions}
\subfigure[]{\includegraphics[width=0.43\textwidth]{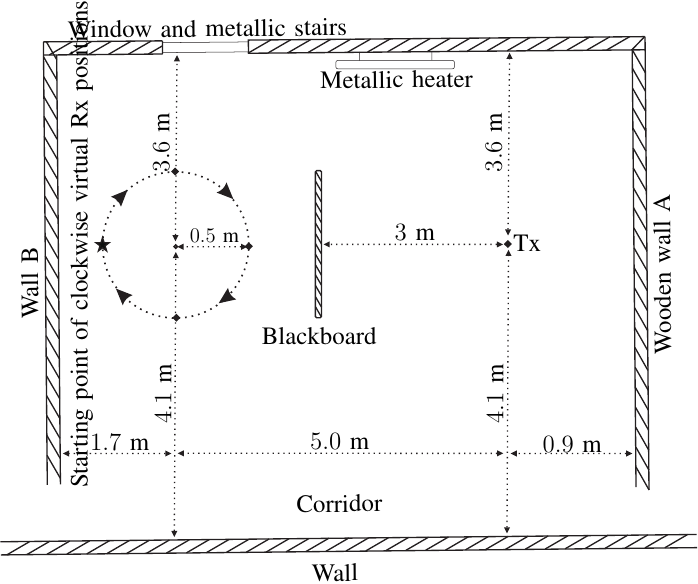}\label{fig:sketch}}
\psfrag{O}[c][c][0.7]{{\white Virtual UCA}}
\psfrag{P}[l][l][0.7]{{\white Window and metallic stairs}}
\psfrag{Q}[c][c][0.7]{{\white Metallic heater}}
\psfrag{R}[c][c][0.7]{{\white Tx}}
\psfrag{M}[c][c][0.6]{{\white Blackboard}}
\subfigure[]{\includegraphics[width=0.43\textwidth]{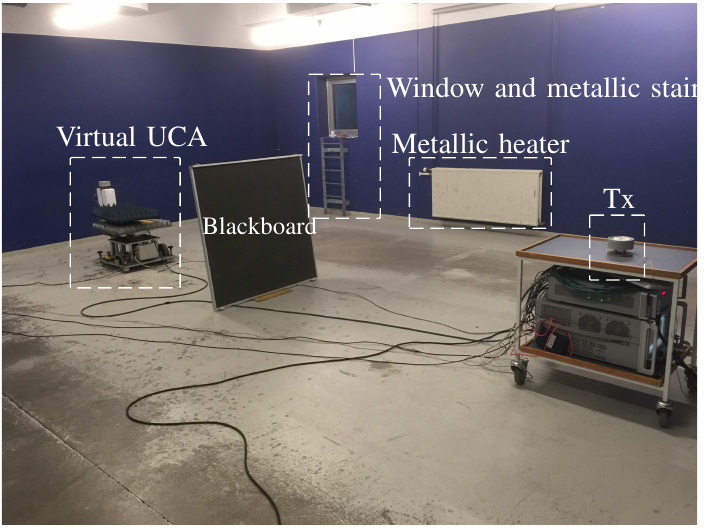}\label{fig:photo}}
\end{center}
\caption{The basement where the measurement campaign was conducted. (a) The top-view sketch of the basement. (b) A photo taken during the OLoS measurement. \label{fig:typical_pdp}}
\end{figure}

Fig.\ \ref{fig:sketch} illustrates the top-view sketch of the basement where the measurement campaign was conducted. The floor space was around $7.7*7.9$ m$^2$ with few objects including one metallic stairs and one metallic heater. Fig.\ \ref{fig:photo} illustrates a photo taken during the measurement. A VNA was used in the channel measurement. An {omnidirectional} biconical antenna was exploited as the Tx and was put on the trolley with the VNA onboard. The height of the Tx to the ground was 0.84 m. An identical biconical antenna was fixed on the turntable as the Rx. A virtual UCA was formed by rotating the Rx clockwise with radius of 0.5 m with 720 steps, i.e. $r=0.5$ and $P=720$.\footnote{The Fraunhofer distance, i.e. $\frac{2\cdot(2r)^2}{\lambda}$, is calculated as 200\,m. Plane wave propagation is considered valid only when the source distance $d_\ell$ is far greater than it.} The distance space between two neighboring Rx positions was 4.4 mm, which is less than the half wavelength (5 mm) at 30\,GHz and hence avoids the spatial aliasing \cite{arraybook}. The Rx height was kept the same with that of the Tx, and the starting point of the virtual UCA was marked as the star in Fig.\ \ref{fig:sketch}, which means that the azimuth and elevation angles of the line of sight (LoS) path were 180$^\circ$ and 90$^\circ$, respectively. Moreover, the distance between the Tx and the UCA center was 5 m.

The frequency range $f$ was set from 28 GHz to 30 GHz with 750 frequency sweeping points, i.e, $K=750$, in the VNA. The frequency step was 2.7 MHz which corresponds to a maximal observable propagation distance of 111 m. It is sufficient to record the propagation paths confined in the basement. The diameter (1\,m) of the UCA is around 6.7 times the intrinsic distance resolution (15 cm) determined by the system bandwidth of 2 GHz. Therefore, the narrowband assumption \cite{526899} or the so-called small-scale characterization \cite{868476} is invalid in the ultra-wideband measurement campaign. Moreover, the Fraunhofer far-filed distance calculated for the UCA is around 200\,m. Thus the near-filed (i.e. spherical wave) propagation must be taken into account.

Two different scenarios, i.e., LoS scenario and Obstructed-LoS (OLoS) scenario, were considered during the measurement. In the OLoS scenario, a blackboard with a metallic substrate and dimension of $1.2*1.2$ m$^2$ was placed in between the Tx and the UCA to block the paths in LoS direction. Furthermore, to obtain references for the measured channels in the two scenarios, we also exploited the horn antenna to replace the Rx biconical antenna to repeat the measurements. It is worthy noting that except that the horn antenna was fixed at the UCA center, i.e. $r=0$, the other settings and measurement procedure were kept the same for the reference measurements. The specifications of the measurement campaign and of the antennas applied in the measurement campaign are included in Table \ref{table:measspec} and Table \ref{table:antespec}, respectively. The presented antenna gains and half-power beamwidths in Table \ref{table:antespec} are the values evaluated in the considered frequency range 28-30\,GHz. Readers may refer to \cite{Fan2016Measured} for the detailed description of the measurement system including the calibration, etc.

\begin{table}
\centering
\caption{Measurement specifications applied in the measurement campaign}
\scalebox{1}{
\begin{tabular}{lclc}
\hline
\multicolumn{4}{c}{{{{\textit{Measurement specifications}}}}}\\\hline
Tx and Rx Heights & 0.84 m &  Frequency range & 28-30 GHz     \\
Radius of bicoinal array & 0.5 m & Frequency points & 750  \\
Number of virtual Rxs & 720 & LoS distance & 5 m \\ \hline
\end{tabular}}
\label{table:measspec}
\end{table}

\begin{table}
\centering
\caption{Specifications of antennas applied in the measurement campaigns}
\scalebox{1}{
\begin{tabular}{lcc}
\hline
\multicolumn{3}{c}{{{{\textit{Antenna specifications}}}}}\\\hline
 Antenna type             & Biconical  & Reference horn     \\
Operating frequency range & 2-30 GHz   & 26.4-40.1 GHz   \\
Gain                      & 6 dBi      &  19 dBi   \\
{HPBW in azimuth}      & Omni       & 20$^\circ$ \\
Polarization              & Vertical   & Vertical \\
\hline
\end{tabular}}
\vspace{2mm}
\\
HPBW: Half power beamwidth
\label{table:antespec}
\end{table}

\subsection{Channel observations}\label{section:channel_characterization}

In this section, we present the measurement results to shed lights on the physical propagation mechanisms for the ultra-wideband large-scale array system.
\begin{figure}
\begin{center}
\psfrag{A}[c][c][0.75]{Propagation delay [s]}
\psfrag{B}[c][c][0.75][90]{Virtual Rx index ($p$)}
\psfrag{C}[c][c][0.75][90]{Received power [dB]}
\psfrag{L}[l][l][0.6]{{\white Trajectory $2$, Wall B}}
\psfrag{M}[l][l][0.6]{{\white Trajectory $1$, Wall A}}
\psfrag{N}[r][r][0.6]{{\white LoS}}
\subfigure[]{\includegraphics[width=0.45\textwidth]{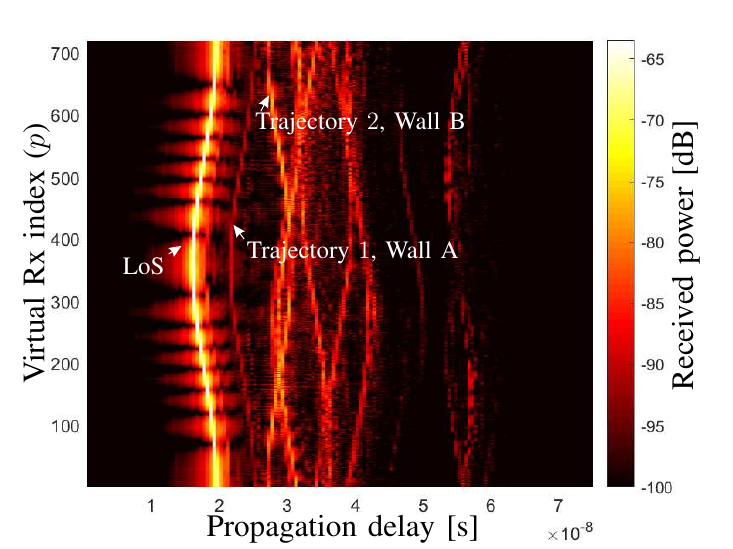}\label{fig:lospdp}}
\subfigure[]{\includegraphics[width=0.45\textwidth]{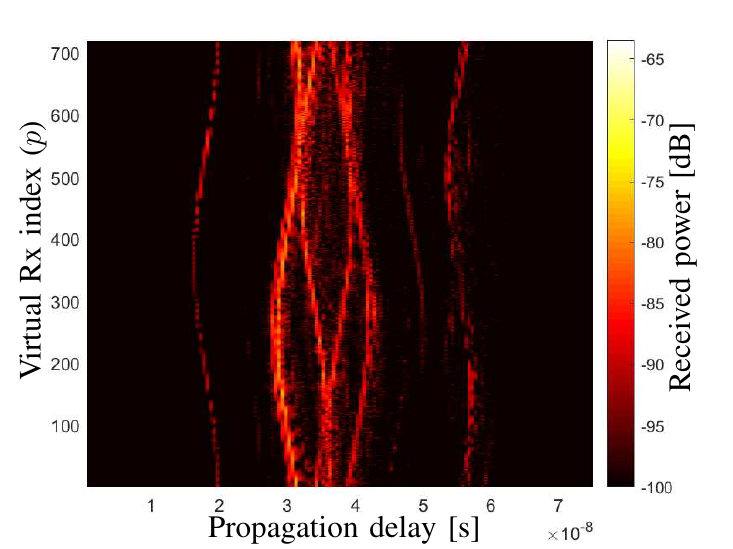}\label{fig:nlospdp}}
\end{center}
\caption{CPDPs obtained by IDFT across the virtual Rxs. (a) LoS scenario. (b) OLoS scenario. \label{fig:cpdps}}
\end{figure}
By applying the inverse discrete Fourier transform (IDFT) to the array output $Y(p,f)$ with respect to $f$, the channel impulse responses (CIRs) $h(p,\tau)$ across the $P$ antennas can be obtained. Here, we omit $\Theta$ for a compact notation of $Y(p,f;\Theta)$. Fig.\ \ref{fig:lospdp} and Fig.\ \ref{fig:nlospdp} illustrate the measured concatenated power delay profiles (CPDPs), i.e., $\left| h(p,\tau) \right|^2$ for the LoS scenario and OLoS scenario, respectively. The further delay range is not shown as no paths are present with the selected dynamic range. By observing and comparing Fig.\ \ref{fig:lospdp} and Fig.\ \ref{fig:nlospdp}, we have the following observations.
\begin{itemize}
\item Finite trajectories exist in both Fig.\ \ref{fig:lospdp} and Fig.\ \ref{fig:nlospdp}. 
    The delay shifts of one path across virtual Rxs can be clearly observed. However, limited by the IDFT operation, sidelobes along these trajectories are also obvious, e.g., along the LoS path as illustrated in Fig.\ \ref{fig:lospdp}.

\item The shapes of trajectories vary due to their different angles-of-arrivals (AoAs). For example, the LoS path in Fig.\ \ref{fig:lospdp} is dominant with a ``(''-alike trajectory. This is consistent with the measurement setup that the virtual UCA elements started at the furthest position to the Tx. It can be expected that these trajectories become straight lines in the narrowband system.

\item  The spatial non-stationarity of the trajectories across the array elements can be clearly observed. That is, one propagation path may present ``birth-death'' behaviour (or break) across the elements when the array aperture is large. This has also been observed in the other works, e.g. in \cite{7172496,7817797}. However, the non-stationarity has not been considered in the literature regarding the channel parameter estimation to our best knowledge.

\item  In the OLoS scenario as illustrated in Fig.\ \ref{fig:nlospdp}, the LoS power attenuates significantly due to the blockage of the blackboard with a metallic substrate. Furthermore, two trajectories disappear in Fig.\ \ref{fig:nlospdp}. They are marked as ``trajectory 1'' and ``trajectory 2'' in Fig.\ \ref{fig:lospdp}, respectively. It can be inferred that trajectory 1 was contributed by the ``wall A'' as illustrated in Fig.\ \ref{fig:sketch}, since its trajectory shape is the same with that of LoS path, and the delay difference is consistent with the measurement setup. Similarly, trajectory 2 contributed by the wall B with an opposite trajectory shape also disappears.
\end{itemize}

The above observations demonstrate the fact that the parameters (e.g. delays and azimuths) of propagation paths actually can be roughly obtained by observing the CPDPs. This motivates us to come up with the idea that it is possible to distinguish the individual paths by identify their trajectories in the CPDPs figure, so that the channel parameter $\Theta$ can be estimated. However, the IDFT sidelobes blur these paths, resulting in the interference among paths and limiting the resolution. Furthermore, the trajectory shapes of different propagation paths change with respect to the AoAs, which hinders the trajectory identification. To cope with these issues, a novel algorithm is proposed, and its high-resolution ability and low computation cost are also demonstrated.

\section{The proposed algorithm\label{algorithm}}
The proposed algorithm basically include three parts, i.e. obtaining the high resolution estimation results of element-wise delays and amplitudes for the UCA,
trajectory identification using the phase-mode excitation technique and estimating $\Theta_\ell$ according to the maximum likelihood principle.

\subsection{Element-wise high resolution estimation\label{delay_searching}}
Considering the ultra-wideband nature of the measurement,  (\ref{eq:elesage}) can be rewritten as
\begin{equation}
\begin{aligned}
H_{p,\ell}(f) & = \alpha_{p}^{\ell}e^{-j 2\pi f \tau_{p}^{\ell}}
\end{aligned}
\label{eq:elesagee}
\end{equation}
with
\begin{equation}
\begin{aligned}
\alpha_{p}^{\ell} & = \frac{d_\ell}{d_{p,\ell}} \alpha_\ell
\end{aligned}
\end{equation}
and
\begin{equation}
\begin{aligned}
\tau_{p}^{\ell} & = \tau_\ell - \Delta d_{p,\ell}/c
\end{aligned}
\label{eq:trajecotory}
\end{equation} Two parameters, i.e. $\tau_{p}^{\ell}$ and $\alpha_{p}^{\ell}$ are capable to characterize the $\ell$th path when a single (the $p$th) array element is considered. We denote the parameter vector $\Omega_{p}^{\ell}=[\tau_{p}^{\ell}, \alpha_{p}^{\ell}]$, then the $L$ propagation paths impinging at the $p$th antenna element can be characterized by the parameter vector $\Omega_p = [\Omega_p^1, \cdots, \Omega_p^L]$.


It can be observed from Fig.\,\ref{fig:cpdps} that the conventional spectra-based methods such as the IDFT is not capable to obtain the high-resolution estimation of $\Omega_p$. The sidelobes of individual propagation paths exist and interfere with each other, and the weak paths can be buried by the sidelobes of strong paths. Therefore, the SAGE principle \cite{753729} or the EM principle
\cite{emalgorithm} is applied to estimating $\Omega_p$ element-wise. With an ultra-wide system bandwidth, the SAGE algorithm can resolve two paths with very small relative propagation delay. By updating the parameter estimations sequentially and providing the maximum-likelihood estimation results iteratively, it has been proven practically in \cite{753729} that the SAGE algorithm has the ability to resolve two paths with $\Delta \tau \gtrapprox \frac{1}{5B}$. $B$ and $\Delta \tau$ represent the bandwidth and the relative delay between two paths, respectively. In our measurement campaign, paths with propagation distance differences no less than 3\,cm can be well resolved by using the SAGE algorithm.

\begin{figure}
\begin{center}
\psfrag{A}[c][c][0.75]{Propagation delay [s]}
\psfrag{B}[c][c][0.75][90]{Virtual Rx index ($p$)}
\psfrag{C}[c][c][0.75][90]{Received power [dB]}
\includegraphics[width=0.7\textwidth]{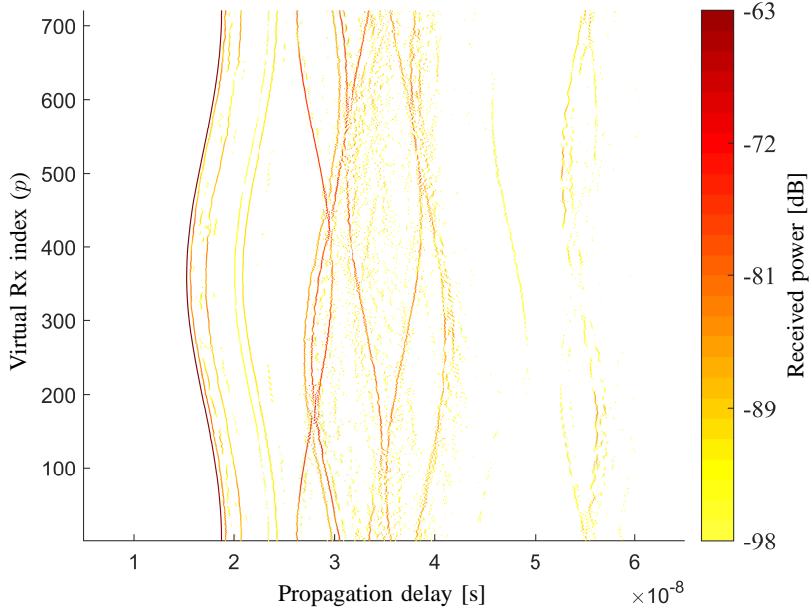}
\end{center}
\caption{UCA element-wise SAGE estimation results $\Omega$. \label{fig:antwisesage}}
\end{figure}

Fig.\ \ref{fig:antwisesage} illustrates the element-wise SAGE estimation results, $\hat{\Omega}=[\hat{\Omega}_1, \cdots, \hat{\Omega}_P]$ for the channel measured in the LoS scenario. In practice, the path number $L$ should be set sufficiently large to fully extract the received power in the channel, which is evaluated as 20 in our case. It can be observed from Fig.\,\ref{fig:antwisesage} that the MPCs are well resolved. Different propagation paths with different trajectories are obvious by visual inspection, e.g. the LoS trajectory is smooth and continuous with the highest power. 
Furthermore, the channel spatial non-stationarity across the array elements can be clearly observed. The path strength along one trajectory may vary, which can result in the break (``birth-death'' behavior) of the trajectory in the extreme case. In the literature, channel stationarity over the array elements is typically assumed for propagation parameter estimation, resulting in unrealistic estimation results and increasing complexity.

\subsection{Trajectory identification based on the phase mode excitation\label{trajectory_identify}}
In this section, we firstly introduce the phase mode excitation technique applied for the 3D wideband spherical propagation scenario, and the trajectory identification for the LoS path is presented.

\subsubsection{Phase mode excitation}
As indicated by (\ref{eq:trajecotory}), $[\tau_\ell, \phi_\ell, \theta_\ell, d_\ell]$ determines the delay trajectory of the $\ell$th propagation path in $\Omega$. The prior knowledge of the parameters is helpful to identify the trajectory.

The phase mode excitation technique \cite{4034131,4533828} have been proposed to transform the spatial-domain array response of the UCA, i.e. $Y(p,f)$ into the phase-mode domain so that a frequency-invariant beamformer can be achieved for the joint delay and angle estimation. However, the frequency-invariant beamformer is effective only for the two-dimensional (2D) propagation scenario, i.e. with all the elevations of paths strictly limited to $90^\circ$.

According to the Taylor expansion, (\ref{eq:distancediff}) can be rewritten as
\begin{equation}
\begin{aligned}
\Delta d_{p,\ell} & = r \sin\theta_\ell\cos(\phi_\ell - \phi_p) + \epsilon_{p,\ell}
\end{aligned}
\end{equation} where $\epsilon_{p,\ell}$ is the remainder of the Taylor series introduced by the spherical wave propagation. To begin with, let us consider the 2D propagation scenario with plane wave assumption, i.e. $\theta_\ell=90^\circ$ and $d_\ell=+\infty$ ($\epsilon_{p,\ell}=0$). The frequency response
$H_{p,\ell}(f)$ contributed by the $\ell$th path at the $p$th Rx defined in (\ref{eq:elesage}) becomes
\begin{equation}
\begin{aligned}
H_{p,\ell}(f) & =  H_{\ell}(f) e^{j 2\pi f \frac{r}{c} \cos(\phi_\ell - \phi_p)} \\
                   & = H_{\ell}(f) \sum_{n=-\infty}^{+\infty} j^n J_n(2\pi f \frac{r}{c})e^{j n(\phi_\ell - \phi_p)}
\end{aligned}
\label{eq:fibf}
\end{equation} The second equality in (\ref{eq:fibf}) holds according to the expansion
\begin{equation}
\begin{aligned}
e^{j\beta\cos\gamma}  = \sum_{n=-\infty}^{+\infty} j^n J_n(\beta) e^{jn\gamma}
\end{aligned}
\end{equation} where $J_n(\cdot)$ is the first kind of Bessel function with order $n$. By involving the so-called phase mode \cite{4533828}, i.e. the basis function $e^{-j m \phi_p}$, and a filter $W_m(f)$, the $m$th phase-mode response is formatted as
\begin{equation}
\begin{aligned}
H_{m,\ell}(f)  &= \frac{1}{P}\sum_{p=0}^{P-1}  H_{p,\ell}(f) e^{-j m \phi_p} W_m(f) \\
               & \hspace{-3mm} = {H_{\ell}(f)} \sum_{n=-\infty}^{+\infty} J_n(2\pi f \frac{r}{c}) W_m(f) e^{-jn\phi_\ell}  \sum_{p=0}^{P-1} \frac{e^{j(m-n)\phi_p}}{P}\\
               & \approx  H_{\ell}(f) J_m(2\pi f \frac{r}{c}) W_m(f) e^{-jm\phi_\ell}\\
               & = H_{\ell}(f) e^{-jm\phi_\ell}
\end{aligned}
\label{eq:fibf1}
\end{equation} The third approximation in (\ref{eq:fibf1}) holds because
\begin{equation}
\begin{aligned}
\frac{1}{P}\sum_{p=0}^{P-1} e^{j(m-n)\phi_p} = 1
\end{aligned}
\label{eq:dd}
\end{equation}
holds only if $n = m + P\cdot z$ where $z$ is an arbitrary integer, and  $J_n(2\pi f \frac{r}{c})$ decrease to near 0 when $z\neq0$. The fourth equability in (\ref{eq:fibf1}) is achieved by making
\begin{equation}
\begin{aligned}
W_m(f)  & = \frac{1}{J_m(2\pi f \frac{r}{c})}
\end{aligned}
\end{equation} The noiseless array response in the phase-mode domain reads
\begin{equation}
\begin{aligned}
Y(m,f;\Theta)  & =  \sum_{\ell=1}^{L} H_{\ell}(f) e^{-jm\phi_\ell}
\end{aligned}
\end{equation} Therefore, the joint delay-azimuth estimation can be accomplished by applying the 2D fast Fourier transform (FFT) to $Y(m,f;\Theta)$.

However, in the realistic channel, it is not practical to assume that all the elevations of paths are exactly $90^\circ$. Moreover, for the large-scale antenna array, spherical wave propagation cannot be neglected. Consequently, the bessel function part in the second step of (\ref{eq:fibf1}) becomes $J_n(2\pi f \frac{r\sin\theta_\ell}{c})$, and the summation part with respect to $p$ in the second step of (\ref{eq:fibf1}) becomes $\sum_{p} e^{j2\pi \frac{\epsilon_{p,\ell}}{c}}  e^{j(m-n)\phi_p}$. The null misalignments between $J_m(2\pi f \frac{r}{c})$ and $J_m(2\pi f \frac{r\sin\theta_\ell}{c})$ destroys the phase-mode transform. The additive phases caused by $\epsilon_{p,\ell}$ may also make (\ref{eq:dd}) invalid. To cope with the 3D propagation scenario with plane wave propagation, the filter
\begin{equation}
\begin{aligned}
G_m(f) = \frac{2}{J_m(2\pi f \frac{r}{c}) + J^{'}_m(2\pi f \frac{r}{c})}
\end{aligned}
\label{eq:newfilter}
\end{equation} was proposed in \cite{7523340}, where $J^{'}_n(\gamma)$ represents the derivative of $J_n(\gamma)$ with respect to $\gamma$. The basic idea is to make use of the cosine-alike oscillation of bessel function
$J_n(\gamma)$ when $\gamma$ is much larger than $n$. Therefore, the denominator of the filter is approximately the envelope of $J_n(\gamma)$. In such a way that the nulls of $J_n(\gamma)$ is avoided, while the amplitude is about 2 times of $J_n(\gamma)$, resulting in the numerator of (\ref{eq:newfilter}) as 2.

\begin{figure}
\begin{center}
\psfrag{R}[c][c][0.75]{Propagation delay [s]}
\psfrag{S}[c][c][0.75][90]{Azimuth [$^\circ$]}
\psfrag{T}[c][c][0.75][90]{Power [dB]}
\psfrag{error}[c][c][0.75]{Delay error [ns]}
\psfrag{elevation}[c][c][0.75]{Elevation [$^\circ$]}
\psfrag{distance}[c][c][0.75]{Source distance [m]}
\psfrag{aerror}[c][c][0.75]{Azimuth error [$^\circ$]}
\subfigure[]{\includegraphics[width=0.45\textwidth]{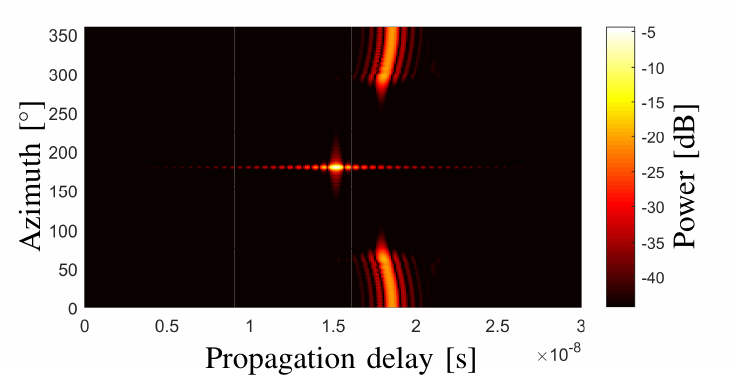}\label{fibfspectrum}}
\subfigure[]{\includegraphics[width=0.45\textwidth]{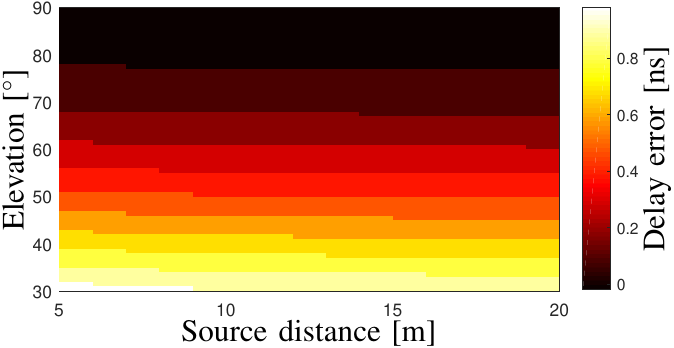}\label{delayerror}}
\subfigure[]{\includegraphics[width=0.45\textwidth]{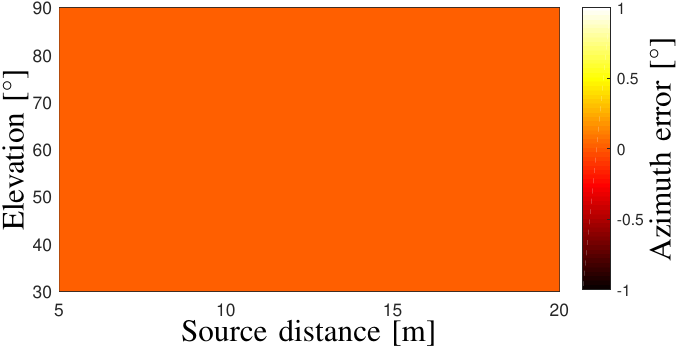}\label{azimutherror}}
\end{center}
\caption{Investigations for the phase mode excitation technique through simulations. (a) power spectrum of a single spherical wave (path). (b) estimation errors of delay. (c) estimation errors of azimuth.\label{fig:fibfspectrumm}}
\end{figure}

Fig.\ \ref{fibfspectrum} illustrates the simulated noiseless delay-azimuth power spectrum $p(\tau,\theta)$ of a single path with
$\Theta = [15\,\text{ns}, 180^\circ, 70^\circ, 5\,\text{m}, 1]$
by applying the filter defined in (\ref{eq:newfilter}). The other simulation configurations are the same with that as applied in the measurement campaign. The estimated delay and azimuth angle indicated by the spectrum maxima in Fig.\ \ref{fibfspectrum} are $15\,\text{ns}$ and $180^\circ$, which are identical to the set values. Note that the elevation angle and source distance cannot be estimated in this step. In addition, it can be observed that sidelobes exist in both delay and azimuth domain, and the power is lower than the real path power. The reasons include that \textit{i)} the null locations and the intervals between these nulls of $J_m(2\pi f \frac{r\sin\theta_\ell}{c})$ shift with respect to that of $J_m(2\pi f \frac{r}{c})$; \textit{ii)} and that the additive spherical-wave phases are introduced. Nevertheless, the peak of the power spectrum is able to indicate the real $\tau$ and $\theta$ of this path. To investigate the effect of elevation and source distance to the estimation results of delay and azimuth, Fig.\,\ref{delayerror} illustrates the estimation errors of the path delay with different source distances and elevations, and Fig.\,\ref{azimutherror} illustrates that of azimuths. It can be observed that for the spherical wave propagations with different elevations and source distances, phase mode excitation works well to obtain rough estimation results of azimuth and delay for the dominant path, with low computation complexity of 2D FFT.


\subsubsection{Trajectory identification\label{trajectory_identification}}

The steps to identify the trajectory for the $\ell$th propagation path in $\hat{\Omega}^{-(\ell-1)}$ based on the phase mode excitation are as follows, where the superscript $-(\ell-1)$ indicates the remainder of $\Omega$ after removing the first $\ell-1$ path trajectories already identified, e.g., $\hat{\Omega}^{0}= \hat{\Omega}$.

\begin{figure}
\begin{center}
\psfrag{A}[c][c][0.75]{Propagation delay [s]}
\psfrag{B}[c][c][0.75][90]{Virtual Rx index ($p$)}
\psfrag{C}[c][c][0.75][90]{Received power [dB]}
\psfrag{d}[l][l][0.7]{$A_{90^\circ}$}
\psfrag{e}[l][l][0.7]{$A_{30^\circ}$}
\psfrag{elevation}[c][c][0.75]{$\theta_1$ [$^\circ$]}
\psfrag{Count}[c][c][0.75]{$C(A_{\theta_1})$}
\subfigure[]{\includegraphics[width=0.45\textwidth]{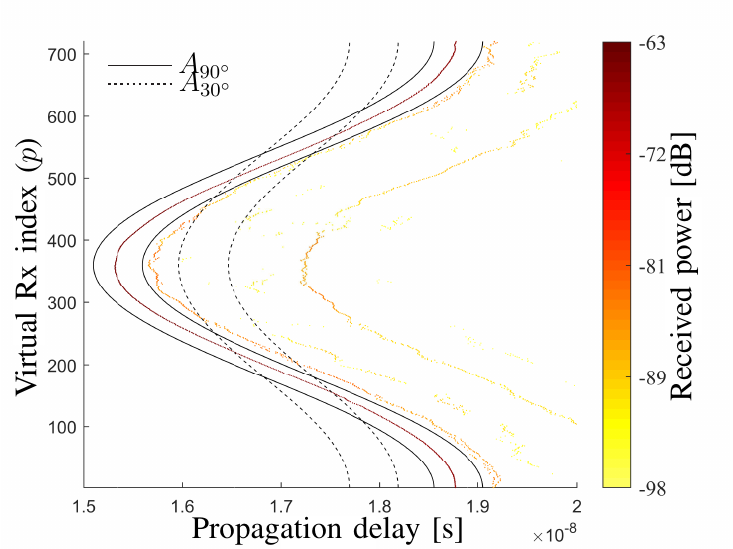}\label{fig:losident}}
\subfigure[]{\includegraphics[width=0.45\textwidth]{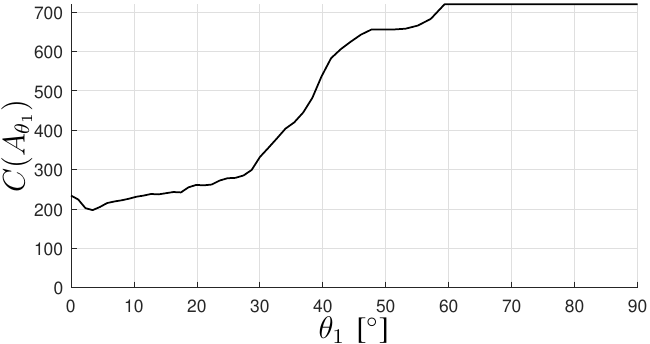}\label{fig:path_area}}
\end{center}
\caption{Identification for the LoS path. (a) LoS path with example $A_{\theta_1}$. (b) $C(A_{\theta_1})$ with respect to different $\theta_1$. \label{fig:losindentification}}
\end{figure}

\begin{enumerate}

\item Reconstruct the current array output $\hat{Y}^{-(\ell-1)}(p,f)$ according to (\ref{eq:elesagee}) using $\hat{\Omega}^{-(\ell-1)}$. 

\item Apply the phase-mode transform to $\hat{Y}^{-(\ell-1)}(p,f)$ and then 2D FFT to $\hat{Y}^{-(\ell-1)}(m,f)$ to obtain the angle-delay power spectrum $\hat{z}^{-(\ell-1)}(\tau,\phi)$. The azimuth $\hat{\phi}_\ell$ and delay $\hat{\tau}_\ell$ of the $\ell$th path are obtained by checking the peak of $\hat{z}^{-(\ell-1)}(\tau,\phi)$.

\item Define the area $A_{\theta_\ell}$ with the trajectory pairs ($T_{\theta_\ell} - \Delta\tau$, $T_{\theta_\ell} + \Delta\tau$), where
\begin{equation}
\begin{aligned}
T_{\theta_\ell} = \hat{\tau}_\ell - \frac{r}{c} \sin\theta_\ell\cos(\hat{\phi}_\ell - \phi_p)
\end{aligned}
\end{equation} Note that $d_\ell$ is ignored to avoid joint 2D searching. $\Delta\tau$ is aimed to tolerate the delay variation caused by the disturbance and the ignorance of $d_\ell$. As an example, Fig.\ \ref{fig:losident} illustrates the path areas $A_{90^\circ}$ and $A_{30^\circ}$ for the LoS path identification, where $\Delta_\tau$ is practically set as $\frac{1}{2B}$. It can be observed that the LoS path trajectory is confined within $A_{90^\circ}$ perfectly however not in $A_{30^\circ}$, indicating the elevation angle is close to 90 degrees.
Hence, the area in which the $\ell$th trajectory is confined can be identified as
\begin{equation}
\begin{aligned}
\hat{\theta}_\ell = \arg \max_{{\theta_\ell}} C(A_{{\theta_\ell}})
\end{aligned}
\label{eq:eleest}
\end{equation} where $C(A_{{\theta_\ell}})$ denotes the number of array elements with propagation paths confined in $A_{\theta_\ell}$. As an example, Fig.\ \ref{fig:path_area} illustrates the $C(A_{{\theta_1}})$ with respect to different $\theta_1$s for the LoS path. It can be observed from Fig.\ \ref{fig:path_area} that $C(A_{{\theta_1}})$ is 720 (i.e. $P$) with elevation angles from $60^\circ$ to $90^\circ$. 
That is, by applying (\ref{eq:eleest}), a rough estimation or a rough initialization of $\theta_\ell$ can be obtained in this step. Note that due to the spatial non-stationarity, the maximum value of $C(A_{{\theta_\ell}})$ can be less then the elements number $P$. Furthermore, it is possible that for one element, more than one path can be contained in $A_{\hat{\theta}_\ell}$. To cope with the issue, the mean power of paths in $A_{\hat{\theta}_\ell}$ is calculated, and the path with minimal power difference from the mean power is selected element-wise. The selected paths forming the trajectory in $\hat{\Omega}^{-(\ell-1)}$ are denoted as $\Omega^{\ell}$.


%

\end{enumerate}

\begin{figure*}
\centering
		 \psfrag{A}[c][c][0.75]{Propagation delay [s]}
         \psfrag{B}[c][c][0.75][90]{Azimuth [$^\circ$]}
         \psfrag{C}[c][c][0.75][90]{Power [dB]}
         \psfrag{D}[c][c][0.75][90]{Virtual Rx index ($p$)}
         \psfrag{E}[c][c][0.75][90]{Power [dB]}
         \psfrag{F}[l][l][0.7]{$\hat{\Omega}^{1}$ in $A_{\hat{\theta}_\ell}$} 
         \psfrag{G}[l][l][0.7]{$\hat{\Omega}^{2}$ in $A_{\hat{\theta}_\ell}$}  
         \psfrag{P}[l][l][0.7]{{\white Dominant peak}}
        \subfigure[$\hat{\Omega}^{0}(\hat{\Omega})$]{\includegraphics[width=0.41\textwidth]{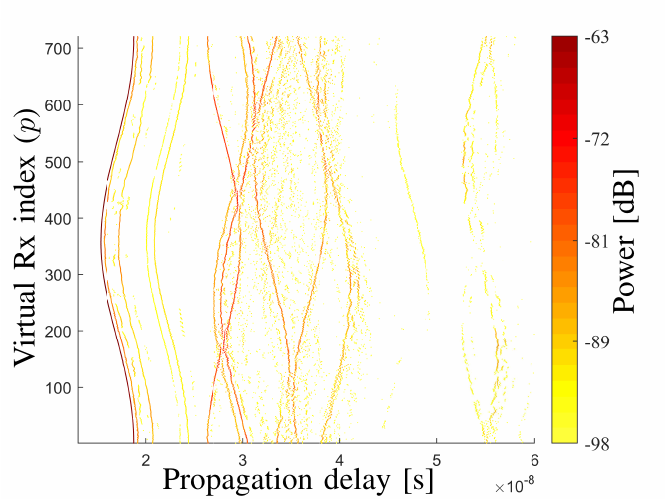}}\hspace{1cm}
        \subfigure[Phase-mode power spectrum of $\hat{\Omega}$]{\includegraphics[width=0.41\textwidth]{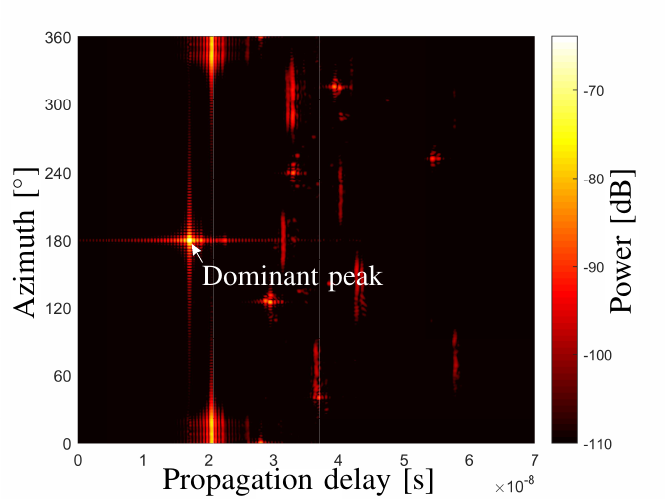}}
        \subfigure[Identification for the first path $\hat{\Omega}^{1}$]{\includegraphics[width=0.41\textwidth]{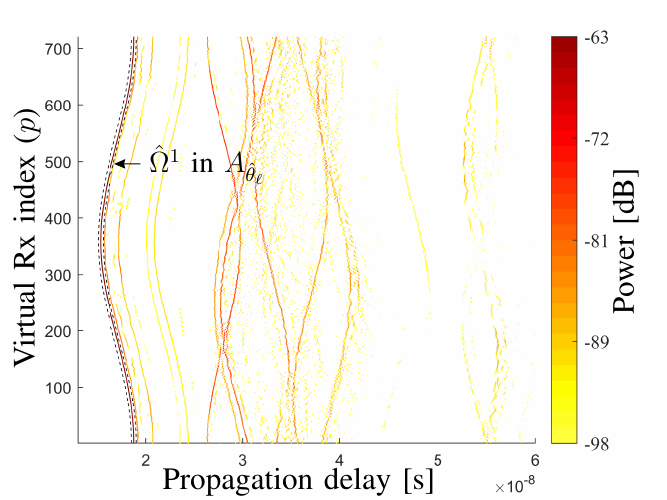}}\hspace{1cm}
        \subfigure[$\hat{\Omega}^{-1}$ obtained by removing $\hat{\Omega}^{1}$ from $\hat{\Omega}$]{\includegraphics[width=0.41\textwidth]{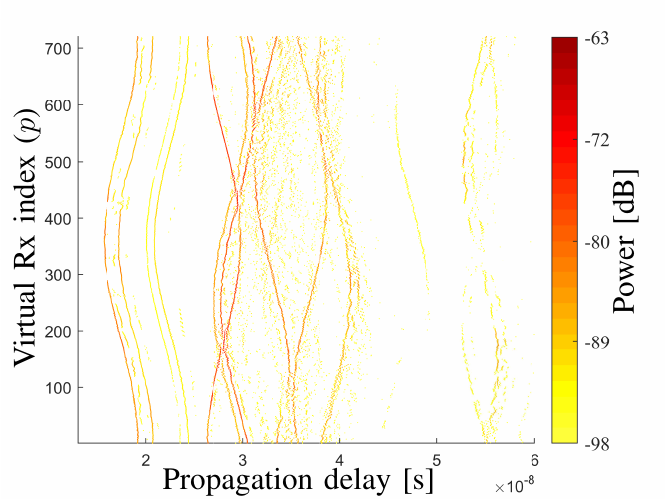}}
        \subfigure[Phase-mode power spectrum of $\hat{\Omega}^{-1}$]{\includegraphics[width=0.41\textwidth]{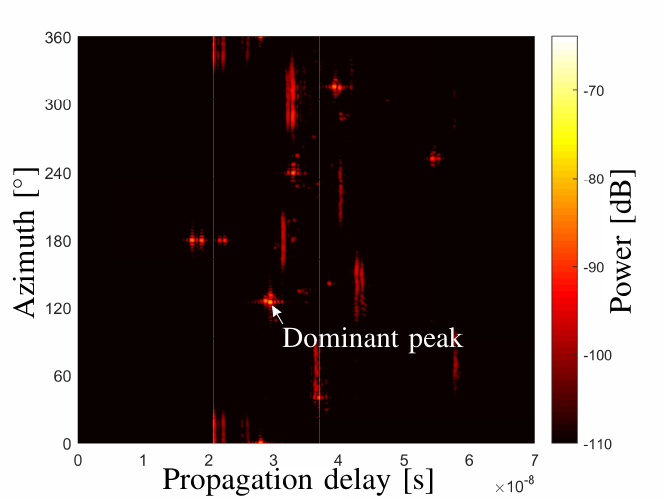}}\hspace{1cm}
        \subfigure[Identification for the second path $\hat{\Omega}^{2}$]{\includegraphics[width=0.41\textwidth]{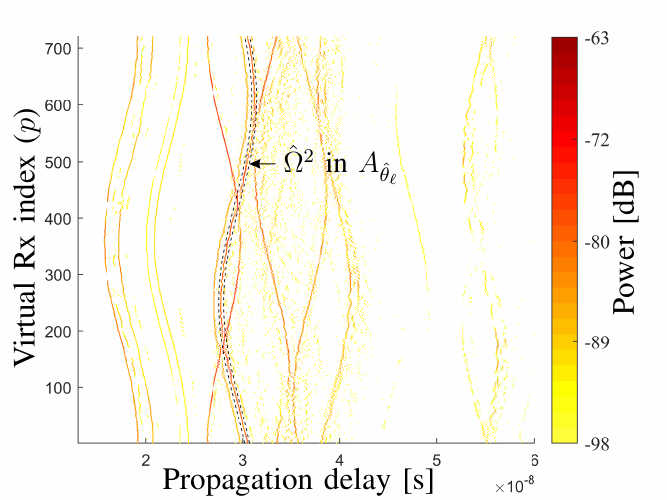}}
	\caption{An example plot of the algorithm implementation procedure for the first two propagation paths.\label{fig:tracking_process}}
\end{figure*}

As an example, readers can refer to the identification steps for the LoS trajectory ($\hat{\Omega}^{1}$) as illustrated in Figs.\,\ref{fig:tracking_process}(a)-\ref{fig:tracking_process}(c).

\subsection{Maximization estimation for $\Theta_\ell$ \label{ME}}
The rough estimation results of delay, azimuth and elevation, i.e., $[\hat{\tau_\ell}, \hat{\phi}_\ell, \hat{\theta}_\ell]$, have been obtained in the former steps. The estimation of $\Theta_\ell$ for the $\ell$th path include three steps as follows. 

\begin{enumerate}

\item Reconstruct the array output $\hat{H}(p,f;\Theta_\ell)$ contributed by this single path according to (\ref{eq:elesagee}) using $\Omega^\ell$ obtained in the trajectory identification.

\item Update the estimation of $[\hat{\phi}_\ell, \hat{\theta}_\ell, \hat{d_\ell}]$ by solving the following maximization problem
\begin{equation}
\begin{aligned}
(\hat{\phi}_\ell, \hat{\theta}_\ell, \hat{d}_\ell) = \arg \max_{{\phi}, {\theta}, d} \text{vec}\{\hat{H}(p,f_k;\Theta_\ell)\}^T  \text{vec}\{W^{*}(p,f_k)\}
\end{aligned}
\end{equation} where $W(p,f_k)$ is generated by using (\ref{eq:elesage}) with parameter set $[0, {\phi}, {\theta}, d, 1]$, $d$ is no larger than $c\hat{\tau}_\ell$, $\text{vec}\{\cdot\}$ denotes the vectorization of the argument matrix, and $(\cdot)^{*}$ represents the conjugation of the argument. Although the maximization problem is solved by 3D searching, the searching space is heavily reduced with known prior information of $\tau_\ell$, $\phi_\ell$ and $\theta_\ell$. Furthermore, without delay considered, only one single frequency point $f_k$ is selected for the maximization, which also decreases the computation load significantly.

\item Update $\hat{\tau}_\ell$ by solving the following maximization problem
\begin{equation}
\hat{\tau}_\ell = \arg \max_{\tau} \text{vec}\{\hat{H}(p,f;\Theta_\ell)\}^T \text{vec}\{W^{*}(p,f)\}
\end{equation} where $W(p,f)$ is generated by using (\ref{eq:elesage}) with parameter set $[\tau, \hat{\phi}_\ell, \hat{\theta}_\ell, \hat{d}_\ell, 1]$. The maximization is achieved by 1D searching with prior information of $\tau_\ell$.

\item Calculate the amplitude $\hat{\alpha}_\ell$ as
\begin{equation}
\hat{\alpha}_\ell = \frac{1}{C K} \text{vec}\{\hat{H}(p,f;\Theta_\ell)\}^T \text{vec}\{W^{*}(p,f)\} \label{eq:nonsta}
\end{equation} where $C$ is the number counted in Sect. \ref{trajectory_identification}, i.e. $C=C(A_{\hat{\theta}_\ell})$, and the $W(p,f)$ here is generated using (\ref{eq:elesage}) with parameter set $[\hat{\tau}_\ell, \hat{\phi}_\ell, \hat{\theta}_\ell, \hat{d}_\ell, 1]$.
\end{enumerate}

\subsection{Algorithm implementation}

The channel parameters of multiple propagation paths are estimated through an ``identification-removing'' operation. The procedure is elaborated in the pseudocodes listed in Algorithm\,1.

\begin{codebox}
\Procname{\textbf {Algorithm\,1: The channel parameter $\Theta$ is estimated according to the procedure described by the following pseudo-codes:}\\ \textbf{Input:} Measured array output $Y(p,f)$\\ \textbf{Output:} The parameter $\Theta$ characterizing the 3D spherical propagation channel}
\li Obtain element-wise high-resolution $\hat{\Omega}$ from $Y(p,f)$;    \label{li:for}
\li Let the path number $\ell$ be 1;
\li Let $\hat{\Theta}$ be an empty set.
\li  \textbf{while True}
\li \quad        \textbf{if} $\ell=1$ \textbf{then}
\li \quad\quad         $\hat{\Omega}^{-(\ell-1)}=\hat{\Omega};$
\li \quad        \textbf{else}
\li \quad\quad         $\hat{\Omega}^{-(\ell-1)}=\hat{\Omega}-\sum_{n=1}^{\ell-1} \hat{\Omega}^{n};$
\li \quad        \textbf{end if}
\li \quad Obtain $\hat{\Omega}^{\ell}$ from $\hat{\Omega}^{-(\ell-1)}$ according to Sect. \ref{trajectory_identification};
\li \quad Obtain $\hat{\Theta}_\ell$ from $\hat{\Omega}^{\ell}$ according to Sect. \ref{ME};
\li \quad        \textbf{if} $C<C_s$ \textbf{then}
\li \quad\quad         \textbf{ break};
\li \quad        \textbf{end}
\li $\hat{\Theta}=\hat{\Theta}+\hat{\Theta}_\ell;$
\li $\ell=\ell+1;$
\li  \textbf{end while}
\end{codebox}

\begin{figure}
\begin{center}
\psfrag{A}[c][c][0.75]{Propagation delay [s]}
\psfrag{B}[c][c][0.75][90]{Virtual Rx index ($p$)}
\includegraphics[width=0.6\textwidth]{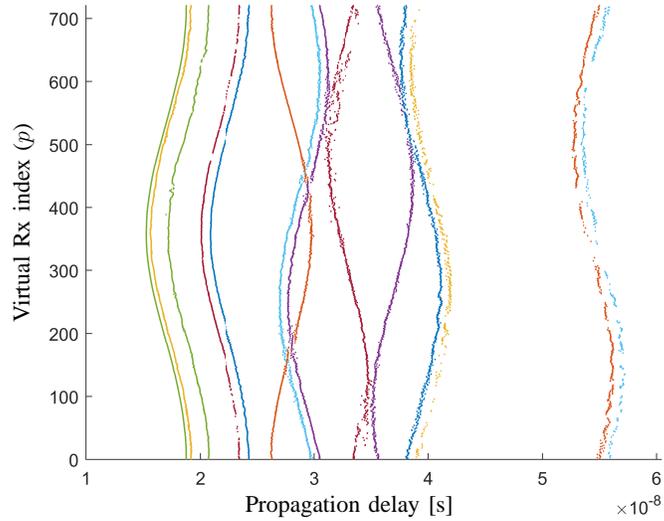}
\end{center}
\caption{Identified trajectories of different propagation paths. Totally 14 paths are estimated in this case. \label{fig:trajectories}}
\end{figure}

Fig.\,\ref{fig:tracking_process} exemplifies the procedure for the first two propagation paths. Note that $C_s$ in line 12 of Algorithm 1 is the threshold for checking the reasonability of $\hat{\Omega}^{\ell}$ identified. The estimation procedure stops if there are not enough elements with propagation paths in the current $\hat{\Omega}^{\ell}$, where it is considered that all the reasonable propagation paths have been identified and estimated. Fig.\ \ref{fig:trajectories} illustrates all the $\hat{\Omega}^\ell$s identified by applying Algorithm\,1 to the measured array output in the LoS scenario as illustrated in Fig.\ \ref{fig:lospdp}. In this case, $C_s$ is practically chosen as 360, the half of the UCA element number $P$. $\Delta\tau$ is practically chosen as $\frac{1}{2B}$.

\section{Results and remarks}\label{resultsremarks}
In this section, the estimation results of the proposed algorithm are firstly presented, followed by the remarks on the proposed algorithm.
\subsection{Estimation results}
\begin{figure}
\begin{center}
\psfrag{Azimuth [Degree]}[c][c][0.75]{Azimuth [$^\circ$]}
\psfrag{Delay [s]}[c][c][0.75]{Propagation delay [s]}
\psfrag{Power [dB]}[c][c][0.75]{Power [dB]}
\psfrag{S}[c][c][0.75][90]{Azimuth [$^\circ$]}
\psfrag{R}[c][c][0.75]{Propagation delay [s]}
\psfrag{T}[c][c][0.75][90]{Power [dB]}
\psfrag{H}[l][l][0.75]{4 stationary paths}
\psfrag{K}[c][c][0.75]{2 stationary paths}
\psfrag{N}[c][c][0.75]{5 stationary paths}
\hspace{0.2cm}\subfigure[]{\includegraphics[width=0.6\textwidth]{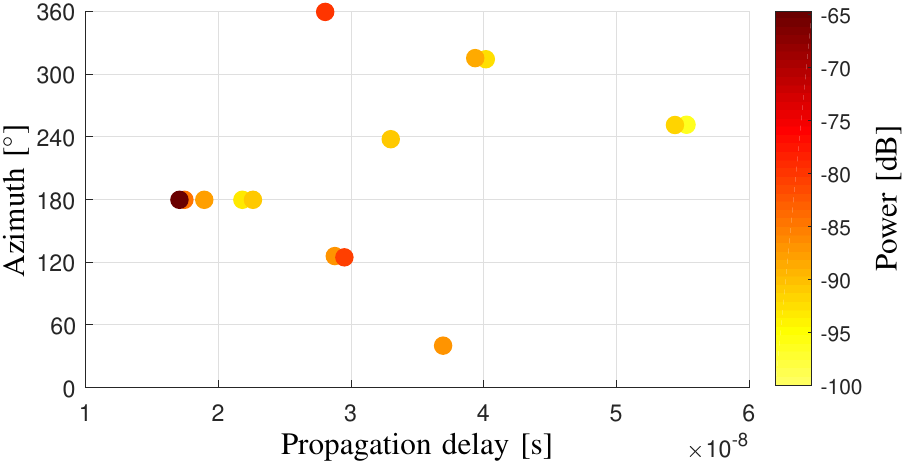}}
\subfigure[]{\includegraphics[width=0.6\textwidth]{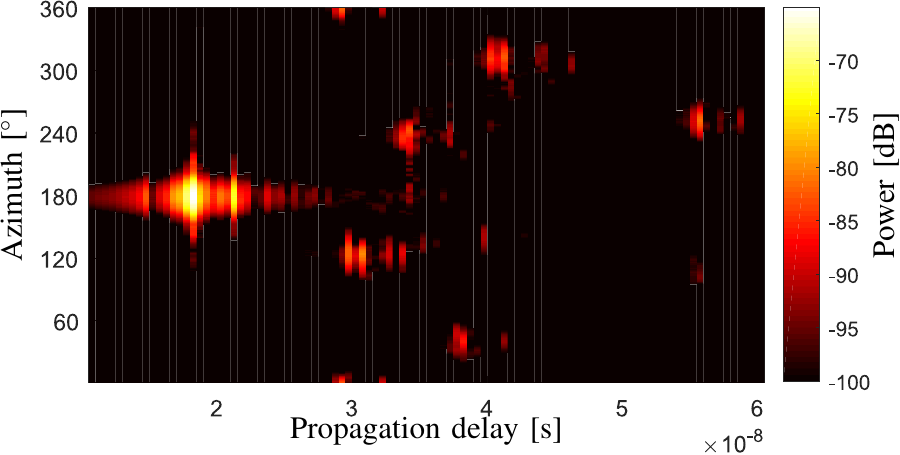}}
\hspace{0.4cm}\subfigure[]{\includegraphics[width=0.6\textwidth]{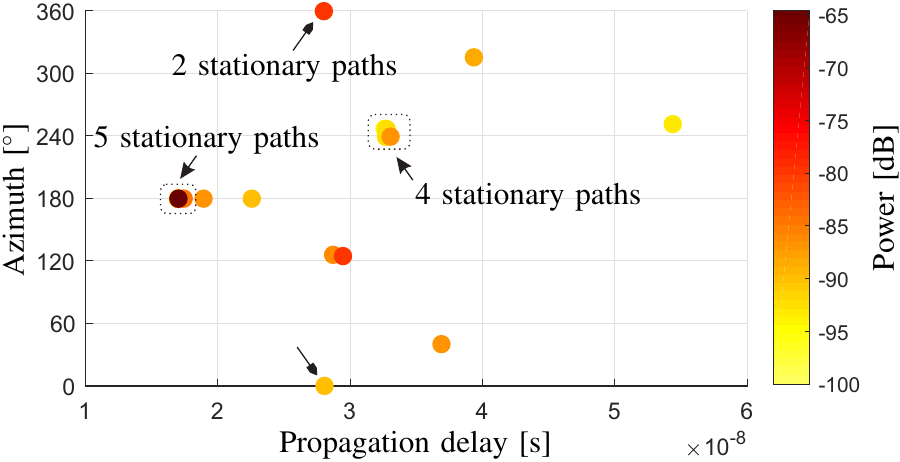}}
\end{center}
\caption{Delay-azimuth power spectra for the LoS scenario. (a) estimation results of the proposed algorithm. (b) measured delay-azimuth power spectrum by using the reference horn antenna. (c) estimated by using the MLE algorithm proposed in \cite{yilin}. \label{fig:delay_azimuth}}
\end{figure}

Fig.\ \ref{fig:delay_azimuth}(a) and Fig.\ \ref{fig:delay_azimuth}(b) illustrate the delay-azimuth power spectra obtained by using the proposed algorithm and measured with the reference horn antenna, respectively, for the LoS scenario. It can be observed that the estimated delay-azimuth power spectrum is in good consistency with the spectrum measured by using the reference horn antenna. Fig.\,\ref{fig:delay_azimuth}(c) also illustrates the estimated delay-azimuth power spectrum by using the MLE algorithm proposed in \cite{yilin} with 4D parameter searchings. By comparing Figs.\,\ref{fig:delay_azimuth}(a) and \ref{fig:delay_azimuth}(c), it can be observed that the two results are very close. However, due to the fact that the channel non-stationarity across the array elements cannot be considered in the MLE algorithm (or the SAGE algorithm), more than one stationary propagation paths (typically with very similar geometrical parameters yet different complex amplitudes) are estimated to mimic the spatial non-stationarity of one propagation path, which results in more artificial paths estimated and increases the estimation complexity.

\begin{figure}
\begin{center}
\psfrag{Azimuth [Degree]}[c][c][0.75]{Azimuth [$^\circ$]}
\psfrag{Delay [s]}[c][c][0.75]{Propagation delay [s]}
\psfrag{Distance [m]}[c][c][0.75]{Source distance [m]}
\psfrag{D}[c][c][0.6]{LoS}
\psfrag{B}[c][c][0.6]{Wall A}
\psfrag{C}[c][c][0.6]{Wall B}
\includegraphics[width=0.6\textwidth]{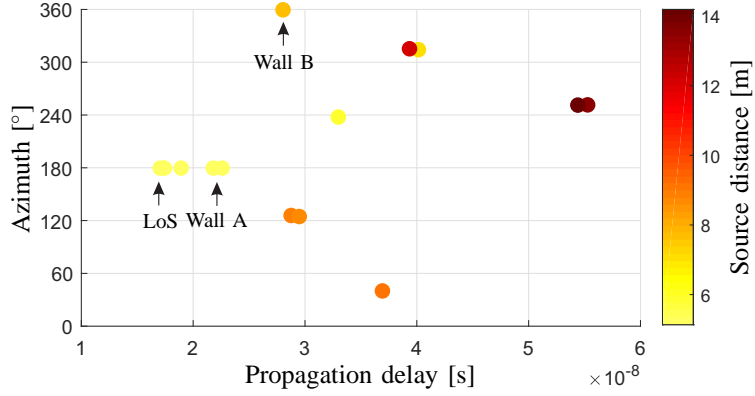}
\end{center}
\caption{Delay-azimuth-distance spectrum estimated for the LoS scenario.\label{fig:delay_distance}}
\end{figure}

Fig.\ \ref{fig:delay_distance} illustrates the source distances for different propagation paths estimated. It can be observed from Fig.\ \ref{fig:delay_distance} that the estimated source distances generally become larger when the propagation distances are larger, which demonstrates the fact that the basement walls were probably smooth and with high reflection effects.

\subsection{Remarks on the proposed algorithm}
The proposed algorithm is essentially a HRPE algorithm based on the maximization-likelihood principle. Compared to the other HRPE algorithms, e.g., in \cite{7981398,7501567,aaltosage,Rimax,yilin}, the proposed algorithms have the following advantages.

 %
%
%
%

\begin{itemize}
\item The proposed algorithm is based on the array-element-wise high resolution estimation in the delay domain as elaborated in Sect. \ref{delay_searching}. In this step, the high resolution ability of the algorithm is guaranteed. Moreover, due to the fact that only delay domain is searched, the complexity in this step is low.
\item In the trajectory identification step as elaborated in Sect. \ref{trajectory_identify}, the proposed algorithm exploits the phase mode excitation technique, the delay trajectories of propagation paths can be identified with 2D FFT with low computation load. Note that the trajectory identification procedure also has the effect of interference cancellation among different paths, which means that the considerable iterations applied in \cite{7981398,7501567,aaltosage,Rimax,yilin} are no longer required. Furthermore, the rough estimation results of delay, azimuth and elevation can be obtained.
\item In the maximization estimation step as elaborated in Sect. \ref{ME}, although a 3D and a 1D parameter searchings are applied, the searching spaces are heavily reduced with the prior initializations obtained in the trajectory identification. Therefore, the computation complexity is decreased significantly compared to the MLE algorithm \cite{yilin} with exhaustive 4D parameter searchings. Moreover, the joint parameter searching avoids the space-alternating operations in the SAGE algorithm \cite{7981398,7501567,aaltosage}, where the violation of OSM can lead to the convergence failure of the SAGE algorithm.
\item  The channel non-stationarity across the array elements is considered in (\ref{eq:nonsta}), which is firstly addressed in the literature for the propagation parameter estimation. The estimation results are more realistic to reproduce the spherical propagation compared to the SAGE \cite{7981398,7501567,aaltosage} and MLE \cite{yilin} algorithms where stationarity paths are assumed.
\end{itemize}



\section{Conclusions} \label{section:conclusion}

In this contribution, a complexity-efficient high resolution parameter estimation (HRPE) algorithm was proposed for the ultra-wideband large-scale uniform circular array (UCA). The algorithm was demonstrated based on a recently conducted measurement campaign in a basement. The delay trajectories of different propagation paths across the virtual array elements were well observed, which can roughly indicate the channel information. The proposed HRPE algorithm takes advantage of the prior channel information obtained in the trajectories and is capable to involve the spatial non-stationarity of realistic channels. Investigations show that the algorithm can obtain the high-resolution results, i.e., the delays, azimuths, elevations, source distances and complex amplitudes for individual spherical propagation paths in the underlying mm-wave channel. Fast initializations and effective interference cancellations reduce the computation load heavily. Furthermore, the consideration of the channel spatial non-stationarity is helpful to obtain more realistic results and to decrease the algorithm complexity, by avoiding the artificial stationary paths being estimated. In addition, it is noteworthy that the basement walls have significant reflection effects for the mm-wave channel at 28-30\,GHz.

\bibliography{ms.bbl}

\end{document}